\documentclass[runningheads]{llncs}
\usepackage[T1]{fontenc}
\usepackage{graphicx}
\usepackage{subfigure,float,hyperref}
\usepackage{amsmath,amssymb,amsfonts}
\hypersetup{
	colorlinks=true,
	linkcolor=blue,
	filecolor=blue,      
	urlcolor=blue,
	citecolor=blue,
}
\usepackage{longtable}
\usepackage{multirow}
\usepackage{float,array}
\usepackage{makecell}
\usepackage{lipsum}
\usepackage[hang,flushmargin]{footmisc}

\begin{document}
\title{H-DenseFormer: An Efficient Hybrid Densely Connected Transformer for Multimodal Tumor Segmentation}
\titlerunning{H-DenseFormer}

%

\author{Jun Shi\inst{1} \and 
Hongyu Kan\inst{1} \and 
Shulan Ruan\inst{1} \and 
Ziqi Zhu\inst{1} \and 
Minfan Zhao\inst{1} \and 
Liang Qiao\inst{1} \and 
Zhaohui Wang\inst{1} \and 
Hong An\inst{1} \and 
Xudong Xue\inst{2}} 


\authorrunning{J. Shi et al.}

\institute{University of Science and Technology of China, Hefei, China \\
\email{shijun18@mail.ustc.edu.cn, han@ustc.edu.cn} \\
\and Hubei Cancer Hospital, Tongji Medical College, Huazhong University of Science and Technology, Wuhan, China}

\maketitle              

\newcommand\blfootnote[1]{ %
    \begingroup
    \renewcommand\thefootnote{}\footnote{ #1}%
    \addtocounter{footnote}{-1}%
    \endgroup
}

\blfootnote{J. Shi and H. Kan contributed equally. This study was supported by the Fundamental Research Funds for the Central Universities (No.YD2150002001).}

\begin{abstract}
Recently, deep learning methods have been widely used for tumor segmentation of multimodal medical images with promising results. However, most existing methods are limited by insufficient representational ability, specific modality number and high computational complexity. In this paper, we propose a hybrid densely connected network for tumor segmentation, named \textbf{H-DenseFormer}, which combines the representational power of the Convolutional Neural Network (CNN) and the Transformer structures. Specifically, H-DenseFormer integrates a Transformer-based Multi-path Parallel Embedding (\textbf{MPE}) module that can take an arbitrary number of modalities as input to extract the fusion features from different modalities. Then, the multimodal fusion features are delivered to different levels of the encoder to enhance multimodal learning representation. Besides, we design a lightweight Densely Connected Transformer (\textbf{DCT}) block to replace the standard Transformer block, thus significantly reducing computational complexity. We conduct extensive experiments on two public multimodal datasets, HECKTOR21 and PI-CAI22. The experimental results show that our proposed method outperforms the existing state-of-the-art methods while having lower computational complexity. The source code is available at \url{https://github.com/shijun18/H-DenseFormer}.

\end{abstract}

\section{Introduction}
Accurate tumor segmentation from medical images is essential for quantitative assessment of cancer progression and preoperative treatment planning \cite{chen2019robust}.
Tumor tissues usually present different features in different imaging modalities.
For example, Computed Tomography (CT) and Positron Emission Tomography (PET) are beneficial to represent morphological and metabolic information of tumors, respectively. In clinical practice, multimodal registered images, such as PET-CT images and Magnetic Resonance (MR) images with different sequences, are often utilized to delineate tumors to improve accuracy. However, manual delineation is time-consuming and error-prone, with a low inter-professional agreement \cite{foster2014review}. These have prompted the demand for intelligent applications that can automatically segment tumors from multimodal images to optimize clinical procedures.

Recently, multimodal tumor segmentation has attracted the interest of many researchers. With the emergence of multimodal datasets (e.g., BRATS~\cite{brats} and HECKTOR \cite{hecktor}), various deep-learning-based multimodal image segmentation methods have been proposed \cite{chen2019robust,dolz2018hyperdense,rodriguez2018multimodal,fu2021multimodal,saeed2022tmss,wang2021transbts}.
Overall, 
large efforts have been made on effectively fusing image features of different modalities to improve segmentation accuracy.
According to the way of feature fusion, the existing methods can be roughly divided into three categories \cite{zhou2019review,guo2019deep}: \emph{input-level fusion}, \emph{decision-level fusion}, and \emph{layer-level fusion}.
As a typical approach, input-level fusion \cite{pereira2016brain,cui2018automatic,kamnitsas2017efficient,zhao2018deep,wang2021transbts} refers to concatenating multimodal images in the channel dimension as network input during the data processing or augmentation stage. This approach is suitable for most existing end-to-end models~\cite{xiao2018resunet,chen2018deeplabv3+}, such as U-Net \cite{ronneberger2015unet} and U-Net++ \cite{zhou2019unet++}.
However, the shallow fusion entangles the low-level features from different modalities, preventing the effective extraction of high-level semantics and resulting in limited performance gains.
In contrast, \cite{zhong20183d} and \cite{kamnitsas2018ensembles} propose a solution based on decision-level fusion.
The core idea is to train an independent segmentation network for each data modality and fuse the results in a specific way.
These approaches can bring much extra computation at the same time, as the number of networks is positively correlated with the number of modalities.
As a compromise alternative, layer-level fusion methods such as HyperDense-Net \cite{dolz2018hyperdense} advocate the cross-fusion of the multimodal features in the middle layer of the network.

In addition to the progress on the fusion of multimodal features,
improving the model representation ability is also an effective way to boost segmentation performance. 
In the past few years, Transformer structure \cite{vaswani2017attention,dosovitskiy2020image,liu2021swin}, centered on the multi-head attention mechanism, has been introduced to multimodal image segmentation tasks. Extensive studies \cite{gao2021utnet,chen2021transunet,hatamizadeh2022unetr,cao2023swin} have shown that the Transformer can effectively model global context to enhance semantic representations and facilitate pixel-level prediction. Wang et al. \cite{wang2021transbts} proposed TransBTS, a form of input-level fusion with a U-like structure, to segment brain tumors from multimodal MR images. TransBTS employs the Transformer as a bottleneck layer to wrap the features generated by the encoder, outperforming the traditional end-to-end models. Saeed et al. \cite{saeed2022tmss} adopted a similar structure in which the Transformer serves as the encoder rather than a wrapper, also achieving promising performance. Other works like \cite{dobko2022combining} and \cite{zhang2022mmformer}, which combine the Transformer with the multimodal feature fusion approaches mentioned above, further demonstrate the potential of this idea for multimodal tumor segmentation.


Although remarkable performance has been accomplished with these efforts, there still exist several challenges to be resolved.
Most existing methods are either limited to specific modality numbers due to the design of asymmetric connections or suffer from large computational complexity because of the huge amount of model parameters.
Therefore, how to improve model ability while ensuring computational efficiency is the main focus of this paper.

To this end, we propose an efficient multimodal tumor segmentation solution named Hybrid Densely Connected Network (\textbf{H-DenseFormer}). First, our method leverages Transformer to enhance the global contextual information of different modalities. Second, H-DenseFormer integrates a Transformer-based Multi-path Parallel Embedding (\textbf{MPE}) module, which can extract and fuse multimodal image features as a complement to naive input-level fusion structure. Specifically, MPE assigns an independent encoding path to each modality, then merges the semantic features of all paths and feeds them to the encoder of the segmentation network. This decouples the feature representations of different modalities while relaxing the input constraint on the specific number of modalities. Finally,  we design a lightweight, Densely Connected Transformer (\textbf{DCT}) module to replace the standard Transformer to ensure performance and computational efficiency. Extensive experimental results on two publicly available datasets demonstrate the effectiveness of our proposed method.





    

\section{Method}

\subsection{Overall Architecture of H-DenseFormer}

\begin{figure*}[]
	\centering
	\includegraphics[width=0.88\columnwidth]{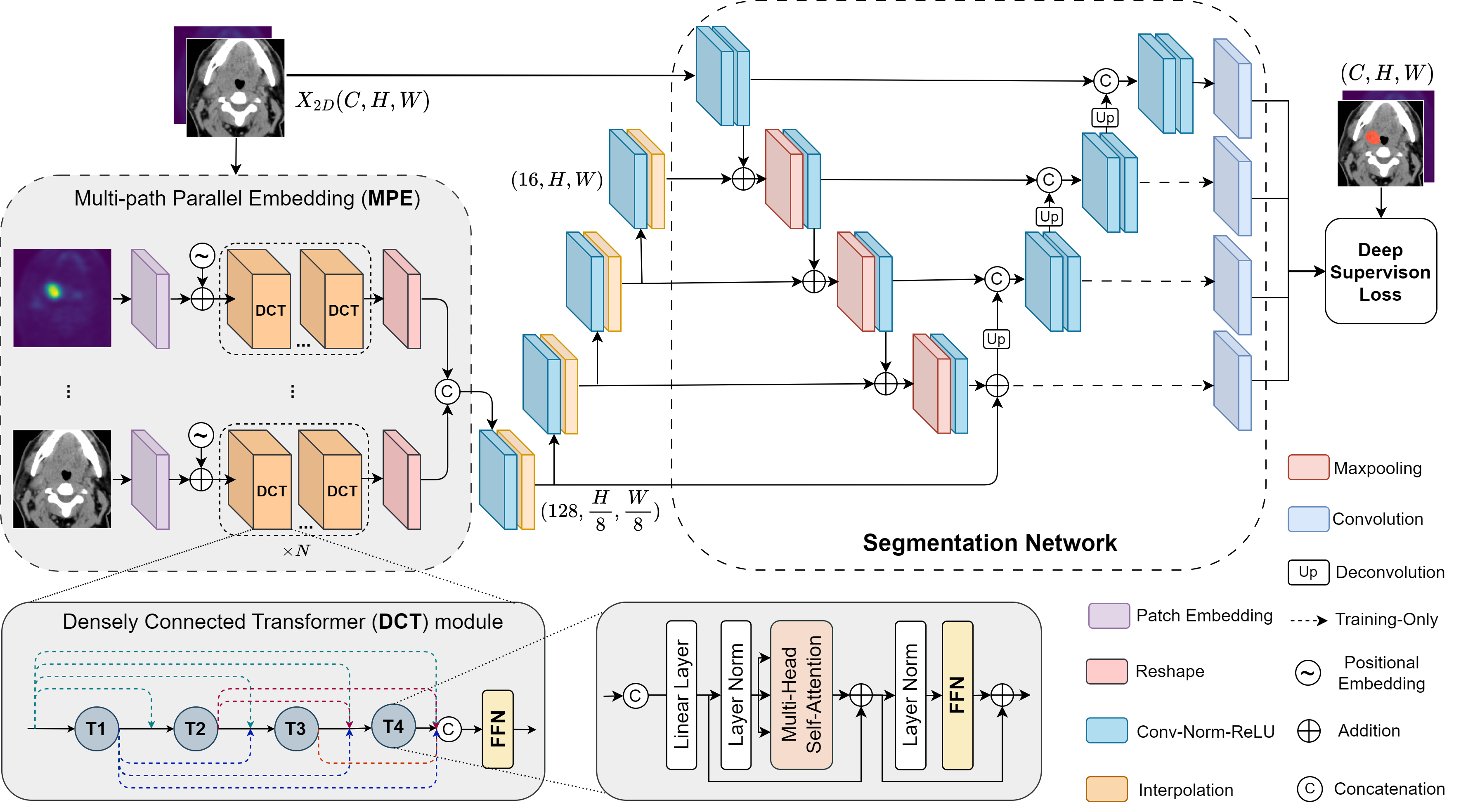} 
	\caption{Overall architecture of our proposed H-DenseFormer.}
	\label{arch-h-denseformer}
\end{figure*}

Fig. \ref{arch-h-denseformer} illustrates the overall architecture of our method. H-DenseFormer comprises a Multi-path Parallel Embedding (MPE) module and a U-shaped segmentation backbone network in form of input-level fusion. The former serves as the auxiliary extractor of multimodal fusion features, while the latter is used to generate predictions. Specifically, given a multimodal image input  $\mathbf{X}_{3D} \in \mathbb{R}^{C\times H \times W \times D}$ or $\mathbf{X}_{2D} \in \mathbb{R}^{C\times H \times W}$ with a spatial resolution of $H \times W$, the depth dimension of D (number of slices) and $C$ channels (number of modalities), we first utilize MPE to extract and fuse multimodal image features. Then, the obtained features are progressively upsampled and delivered to the encoder of the segmentation network to enhance the semantic representation. Finally, the segmentation network generates multi-scale outputs, which are used to calculate deep supervision loss as the optimization target.

\subsection{Multi-path Parallel Embedding}

Many methods \cite{dolz2018hyperdense,guo2019deep,chen2018mri} have proved that decoupling the feature representation of different modalities facilitates the extraction of high-quality multimodal features. Inspired by this, we design a Multip-path Parallel Embedding (MPE) module to enhance the representational ability of the network. As shown in Fig. \ref{arch-h-denseformer}, each modality has an independent encoding path consisting of a patch embedding module, stacked Densely Connected Transformer (DCT) modules, and a reshape operation. The independence of the different paths allows MPE to handle an 
\textbf{arbitrary} number of input modalities. Besides, the introduction of the Transformer provides the ability to model global contextual information. Given the input $X_{3D}$,  after convolutional embedding and tokenization,  the obtained feature of the $i$-th path is $\mathbf{F}_{i} \in \mathbb{R}^{l \times \frac{H}{p} \times \frac{W}{p} \times \frac{D}{p}}$,  where $i \in [1,2,...,C]$ ,   $p=16$ and $l=128$ denote the path size and embedding feature length respectively.  First, we concatenate the features of all modalities and entangle them using a convolution operation. Then, interpolation upsampling is performed to obtain the multimodal fusion feature $\mathbf{F}_{out} \in \mathbb{R}^{k \times \frac{H}{8} \times \frac{W}{8} \times \frac{D}{8}}$, where $k=128$ refers to the channel dimension. Finally, $\mathbf{F}_{out} $ is progressively upsampled to multiple scales and delivered to different encoder stages to enhance the learned representation.

\subsection{Densely Connected Transformer}

Standard Transformer structures \cite{dosovitskiy2020image} typically consist of dense linear layers with a computational complexity proportional to the feature dimension. Therefore, integrating the Transformer could lead to a mass of additional computation and memory requirements. Shortening the feature length can effectively reduce computation, but it also weakens the representation capability meanwhile. To address this problem, we propose the Densely Connected Transformer (DCT) module inspired by DenseNet \cite{huang2017densely} to balance computational cost and representation capability. Fig. \ref{arch-h-denseformer} details the DCT module, which consists of \textbf{four} Transformer layers and a feedforward layer. Each Transformer layer has a linear projection layer that reduces the input feature dimension to $g=32$ to save computation. Different Transformer layers are connected densely to preserve representational power with lower feature dimensions. The feedforward layer at the end generates the fusion features of the different layers. Specifically, the output $\mathbf{z}_j$ of the $j$-th ($j \in [1,2,3,4] $) Transformer layer can be calculated by: 

\begin{equation}
\mathbf{\tilde{z}}_{j-1} = p(cat([\mathbf{z}_0;\mathbf{z}_1;...;\mathbf{z}_{j-1}])),
\end{equation}
\begin{equation}
\mathbf{\tilde{z}}_j = att(norm(\mathbf{\tilde{z}}_{j-1})) + \mathbf{\tilde{z}}_{j-1},
\end{equation}
\begin{equation}
\mathbf{z}_j = f(norm(\mathbf{\tilde{z}}_j)),
\end{equation}
where $\mathbf{z}_0$ represents the original input, $cat(\cdot)$ and $p(\cdot)$ denote the concatenation operator and the linear layer, respectively. The $norm(\cdot)$, $att(\cdot)$, $f(\cdot)$ are the regular layer normalization, multi-head self-attention mechanism, and feedforward layer. The output of DCT is $\mathbf{z}_{out} = f(cat([\mathbf{z}_0;\mathbf{z}_1;... ;\mathbf{z}_{4}]))$. Table \ref{gflops and para} shows that the stacked DCT has lower parameters and computational complexity than a standard Transformer structure with the same number of layers.

\begin{table*}[]
	\caption{Comparison of the computational complexity between the standard 12-layer Transformer structure and the stacked 3 ($=12/4$) DCT modules.}
	\setlength{\tabcolsep}{1mm}
	\centering
	\begin{tabular}{c|c|c|c|c|c}
		\hline
		\multirow{2}{*}{Feature Dimension} & \multirow{2}{*}{Resolution} & \multicolumn{2}{c|}{Transformer} & \multicolumn{2}{c}{Stacked DCT ($\times 3$)} \\  [2pt] \cline{3-6}
		&  & GFLOPs $\downarrow$   & Params $\downarrow$  & GFLOPs $\downarrow$   & Params $\downarrow$  \\ [2pt] \hline
             256  & (512,512) & 6.837    & 6.382M    & \textbf{2.671}  & \textbf{1.435M}  \\ [2pt] \hline
             512  & (512,512) & 26.256   & 25.347M   & \textbf{3.544}  & \textbf{2.290M}  \\ [2pt]\hline
	\end{tabular}
	\label{gflops and para}
\end{table*}

\subsection{Segmentation Backbone Network}

The H-DenseFormer adopts a U-shaped encoder-decoder structure as its backbone. As shown in Fig. \ref{arch-h-denseformer}, the encoder extracts features and reduces their resolution progressively. To preserve more details, we set the maximum downsampling factor to 8. The multi-level multimodal features from MPE are fused in a bitwise addition way to enrich the semantic information. The decoder is used to restore the resolution of the features, consisting of deconvolutional and convolutional layers with skip connections to the encoder. In particular, we employ Deep Supervision (\textbf{DS}) loss to improve convergence, which means that the multiscale output of the decoder is involved in the final loss computation.

\textbf{Deep Supervision Loss.} During training, the decoder has four outputs; for example, the $i$-th output of 2D H-DenseFormer is $\mathbf{O}^{i} \in \mathbb{R}^{c \times \frac{H}{2^i} \times\frac{W}{2^i}}$,  where $i \in [0,1,2,3]$, and $c=2$ (tumor and background) represents the number of segmentation classes. To mitigate the pixel imbalance problem, we use a combined loss of Focal loss \cite{lin2017focal} and Dice loss as the optimization target, defined as follows:
\begin{equation}
\zeta_{FD} = 1 - {\frac{2\sum_{t=1}^{N}{p_tq_t}}{\sum_{t=1}^{N}{p_t+q_t}}} + \frac{1}{N}\sum_{t=1}^{N}{-(1-p_t)^{\gamma}log(p_t)},
\end{equation}
where N refers to the total number of pixels, $p_t$ and $q_t$ denote the predicted probability and ground truth of the $t$-th pixel, respectively, and $r=2$ is the modulation factor. Thus, DS loss can be calculated as follows:
\begin{equation}
\mathbf{\zeta_{DS}} = \sum\alpha_i \cdot \zeta_{FD}(\mathbf{O}^{i},\mathbf{G}^{i}), \alpha_i = 2^{-i}.
\end{equation}
Where $\mathbf{G}^i$ represents the ground truth after resizing and has the same size as $\mathbf{O}^i$. $\alpha$ is a weighting factor to control the proportion of loss corresponding to the output at different scales. This approach can improve the convergence speed and performance of the network.

\section{Experiments}

\subsection{Dataset and Metrics}

To validate the effectiveness of our proposed method, we performed extensive experiments on \textbf{HECKTOR21} \cite{hecktor} and \textbf{PI-CAI22} \footnote[1]{\url{https://pi-cai.grand-challenge.org/}}. HECKTOR21 is a dual-modality dataset for head and neck tumor segmentation, containing 224 PET-CT image pairs. Each PET-CT pair is registered and cropped to a fixed size of (144,144,144). PI-CAI22 provides multimodal MR images of 220 patients with prostate cancer, including T2-Weighted imaging (T2W), high b-value Diffusion-Weighted imaging (DWI), and Apparent Diffusion Coefficient (ADC) maps. After standard resampling and center cropping, all images have a size of (24,384,384). We randomly select 180 samples for each dataset as the training set and the rest as the independent test set (44 cases for HECKTOR21 and 40 cases for PI-CAI22). Specifically, the training set is further randomly divided into five folds for cross-validation. For quantitative analysis, we use the Dice Similarity Coefficient (\textbf{DSC}), the Jaccard Index (\textbf{JI}), and the 95\% Hausdorff Distance (\textbf{HD95}) as evaluation metrics for segmentation performance. A better segmentation will have a smaller HD95 and larger values for DSC and JI. We also conduct holistic t-tests of the overall performance for our method and all baseline models with the two-tailed p < 0.05.

\subsection{Implementation Details}

We use Pytorch to implement our proposed method and the baselines. For a fair comparison, all models are trained from scratch using two NVIDIA A100 GPUs and all comparison methods are implemented with open-source codes, following their original configurations. In particular, we evaluate the 3D and 2D H-DenseFormer on HECKTOR21 and PI-CAI22, respectively. During the training phase, the Adam optimizer is employed to minimize the loss with an initial learning rate of $10^{-3}$ and a weight decay of $10^{-4}$. We use the PolyLR strategy \cite{nnU-Net} to control the learning rate change. We also use an early stopping strategy with a tolerance of 30 epochs to find the best model within 100 epochs. Online data augmentation, including random rotation and flipping, is performed to alleviate the overfitting problem.

\subsection{Overall Performance}

\begin{table*}[]
    \caption{Comparison with existing methods on independent test set. We show the mean$\pm$std (standard deviation) scores of averaged over the 5 folds.}
    \centering
    \begin{tabular}{c|c|c|c|c|c}
        \hline
        Methods (Year) & Params$\downarrow$ & GFLOPs$\downarrow$ & DSC(\%) $\uparrow$    & HD95(mm) $\downarrow$   & JI(\%) $\uparrow$     \\ [2pt]\hline
        \multicolumn{6}{c}{\textbf{HECKTOR21}, two modalities (CT and PET)} \\ [2pt]\hline
        3D U-Net (2016) \cite{cciccek20163dunet} & 12.95M &  629.07   & 68.8$\pm$1.4    & 14.9$\pm$2.2  & 58.0$\pm$1.4  \\ [2pt]
        UNETR (2022) \cite{hatamizadeh2022unetr} & 95.76M &  282.19   & 59.6$\pm$2.5    & 23.7$\pm$3.4  & 48.2$\pm$2.6  \\ [2pt]
        Iantsen et al. (2021) \cite{iantsen2021squeeze} & 38.66M &  1119.75   & 72.4$\pm$0.8    & 9.6$\pm$1.0  & 60.5$\pm$1.1    \\ [2pt]
        TransBTS (2021) \cite{wang2021transbts} & 30.62M &  372.80   & 64.8$\pm$1.0    & 20.9$\pm$3.9  & 52.9$\pm$1.2    \\ [2pt]
        \textbf{3D H-DenseFormer} & \textbf{3.64M} &  \textbf{242.96}   & \textbf{73.9$\pm$0.5}    & \textbf{8.1$\pm$0.6}  & \textbf{62.5$\pm$0.5}   \\ [2pt] \hline
        \multicolumn{6}{c}{\textbf{PI-CAI22}, three modalities (T2W, DWI and ADC)} \\ [2pt]\hline
        Deeplabv3+ (2018) \cite{chen2018deeplabv3+} & 12.33M &  \textbf{10.35}   & 47.4$\pm$1.9   & 48.4$\pm$14.3  & 35.4$\pm$1.7  \\ [2pt]
        U-Net++ (2019) \cite{zhou2019unet++} & 15.97M &  36.08  & 49.7 $\pm$3.9    & 38.5$\pm$6.7  & 36.9$\pm$3.3    \\ [2pt]
        ITUNet (2022)  \cite{kan2022itunet} & 18.13M &  32.67   & 42.1$\pm$2.3    & 67.6$\pm$10.3  & 31.3$\pm$1.6  \\ [2pt]
        Transunet (2021) \cite{chen2021transunet} & 93.23M &  72.62   & 44.8$\pm$3.0   & 59.3$\pm$14.8  & 33.2$\pm$2.5    \\ [2pt]
        \textbf{2D H-DenseFormer} & \textbf{4.25M} &  31.46   & \textbf{49.9$\pm$1.2}    & \textbf{35.9$\pm$8.2}  & \textbf{37.1$\pm$1.2}   \\ [2pt] \hline
    \end{tabular}
    \label{overall}
\end{table*}

\begin{figure}
	\centering
	\includegraphics[width=0.92\columnwidth]{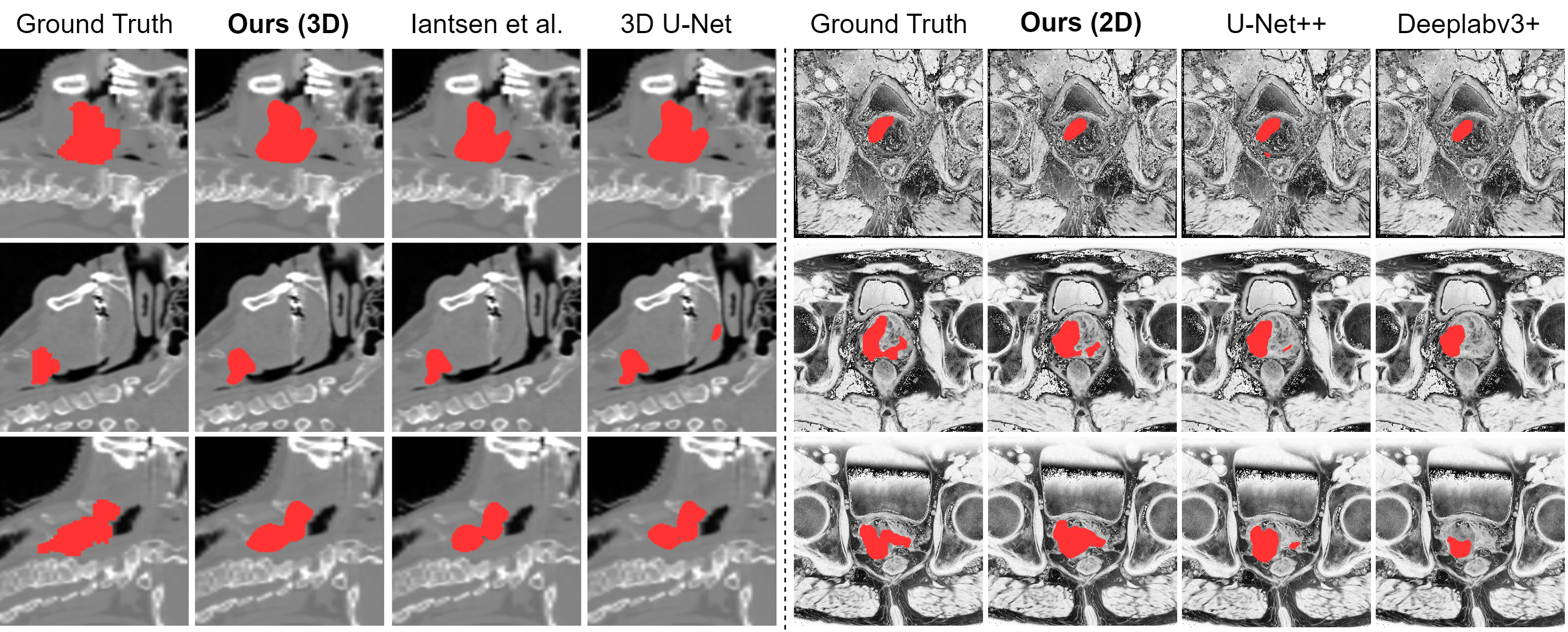} 
	\caption{Visualizations of different models on HECKTOR21 (left) and PI-CAI22 (right).}
	\label{vis}
\end{figure}

Table \ref{overall} compares the performance and computational complexity of our proposed method with the existing state-of-the-art methods on the independent \textbf{test} sets. For HECKTOR21, 3D H-DenseFormer achieves a DSC of 73.9\%, HD95 of 8.1mm, and JI of 62.5\%, which is a significant improvement (p < 0.01) over 3D U-Net \cite{cciccek20163dunet}, UNETR \cite{hatamizadeh2022unetr}, and TransBTS \cite{wang2021transbts}. It is worth noting that the performance of hybrid models such as UNETR is not as good as expected, even worse than 3D U-Net, perhaps due to the small size of the dataset. Moreover, compared to the champion solution of HECKTOR20 proposed by Iantsen et al. \cite{iantsen2021squeeze}, our method has higher accuracy and about \textbf{10} and \textbf{5} times lower amount of network parameters and computational cost, respectively. For PI-CAI22, the 2D variant of H-DenseFormer also outperforms existing methods (p < 0.05), achieving a DSC of 49.9\%, HD95 of 35.9mm, and JI of 37.1\%. Overall, H-DenseFormer reaches an effective balance of performance and computational cost compared to existing CNNs and hybrid structures. For qualitative analysis, we show a visual comparison of the different methods. It is evident from  Fig. \ref{vis} that our approach can describe tumor contours more accurately while providing better segmentation accuracy for small-volume targets. These results further demonstrate the effectiveness of our proposed method in multimodal tumor segmentation tasks.

\subsection{Parameter Sensitivity and Ablation Study}

\begin{table}[h]
    \caption{Parameter sensitivity analysis on DCT depth.}
    \centering
    \begin{tabular}{c|c|c|c|c|c}
        \hline
        DCT Depth & Params$\downarrow$ & GFLOPs$\downarrow$ & DSC(\%) $\uparrow$    & HD95(mm) $\downarrow$   & JI(\%) $\uparrow$     \\ [2pt]\hline
        3    & 3.25M & 242.38 & 73.5$\pm$1.4    & 8.4$\pm$0.7  & 62.2$\pm$1.6  \\ [2pt]
        \textbf{6}    & 3.64M & 242.96 & \textbf{73.9$\pm$0.5}    & \textbf{8.1$\pm$0.6}  & \textbf{62.5$\pm$0.5}    \\ [2pt]
        9   & 4.03M & 243.55 & 72.7$\pm$1.2    & 8.7$\pm$0.6  & 61.2$\pm$1.3  \\ [2pt] \hline
    \end{tabular}
    \label{ab-1}
\end{table}

\begin{table}[]
    \caption{Ablation study of 3D H-DenseFormer, w/o denotes without.}
    \centering
    \begin{tabular}{l|c|c|c}
        \hline
        Method  & DSC(\%) $\uparrow$    & HD95(mm) $\downarrow$   & JI(\%) $\uparrow$     \\ [2pt]\hline
        3D H-DenseFormer \textbf{w/o MPE}  & 72.1$\pm$0.8    & 10.8$\pm$1.1  & 60.4$\pm$0.8 \\
        3D H-DenseFormer \textbf{w/o DCT}  & 70.7$\pm$1.8    & 11.9$\pm$1.9  & 58.6$\pm$2.1 \\
        3D H-DenseFormer \textbf{w/o DS loss} & 72.2$\pm$0.9    & 10.2$\pm$1.0  & 60.1$\pm$1.2 \\
        \textbf{3D H-DenseFormer}  & \textbf{73.9$\pm$0.5}    & \textbf{8.1$\pm$0.6}  & \textbf{62.5$\pm$0.5}   \\ [2pt] \hline
    \end{tabular}
    \label{ab-2}
\end{table}

\textbf{Impact of DCT Depth.} As illustrated in Table \ref{ab-1}, the network performance varies with the change in DCT depth.
H-DenseFormer achieves the best performance at the DCT depth of \textbf{6}. An interesting finding is that although the depth of the DCT has increased from 3 to 9, the performance does not improve or even worsen. We suspect that the reason is over-fitting due to over-parameterization. Therefore, choosing a proper DCT depth is crucial to improve accuracy.

\textbf{Impact of Different Modules.} The above results demonstrate the superiority of our method, but it is unclear which module plays a more critical role in performance improvement. Therefore, we perform ablation experiments on MPE, DCT and DS loss. Specifically, w/o MPE refers to keeping one embedding path, w/o DCT signifies using a standard 12-layer Transformer, and w/o DS loss denotes removing the deep supervision mechanism. As shown in Table \ref{ab-2}, the performance decreases with varying degrees when removing them separately, which means all the modules are critical for H-DenseFormer. We can observe that DCT has a greater impact on overall performance than the others, further demonstrating its effectiveness. In particular, the degradation after removing the MPE also confirms that decoupling the feature expression of different modalities helps obtain higher-quality multimodal features and improve segmentation performance.

\section{Conclusion}
In this paper, we proposed an efficient hybrid model (H-DenseFormer) that combines Transformer and CNN for multimodal tumor segmentation.
Concretely,
a Multi-path Parallel Embedding module and a Densely Connected Transformer block were developed and integrated to balance accuracy and computational complexity.
Extensive experimental results demonstrated the effectiveness and superiority of our proposed H-DenseFormer.
In future work, we will extend our method to more tasks and explore more efficient multimodal feature fusion methods to further improve computational efficiency and segmentation performance. 

\bibliographystyle{splncs04}
\bibliography{paper3097}
\end{document}